\documentclass[pre,twocolumn,showpacs]{revtex4}
\usepackage{epsfig}
\usepackage[]{natbib}
\begin{document}

\title{Fluctuations of time averages for Langevin dynamics in a binding force 
 field}

\author{A. Dechant}
\author{E. Lutz}
\affiliation{Department of Physics,
University of Augsburg,
D-86135 Augsburg, Germany}
\author{D. A. Kessler}
\author{E. Barkai}
\affiliation{Department of Physics, Institute of Nanotechnology and Advanced Materials, Bar Ilan University, Ramat-Gan
52900, Israel}

\begin{abstract}
We derive a simple formula for
the fluctuations of the 
time average $\overline{x}(t)$ around the thermal mean 
$\langle x \rangle_{{\rm eq}}$ for
overdamped Brownian
motion in a binding potential $U(x)$.  
Using a backward Fokker-Planck equation, 
introduced by Szabo, Schulten, and Schulten in the
context of reaction kinetics, we show that 
for ergodic processes these finite measurement time fluctuations
are determined
by the Boltzmann measure. 
For   
the widely applicable logarithmic potential, ergodicity is  broken. 
We quantify the large non-ergodic fluctuations  
and show how they are related to a super-aging 
correlation function. 

\end{abstract}
\maketitle

Current technology permits tracking of trajectories of individual
molecules
 with exquisite precision.
 The motion of a  Brownian particle in a binding
potential field $U(x)$ is used 
to model many such physical,
biological and chemical processes.
 From statistical mechanics, we know
that if the process is ergodic, and if the measurement time $t \to \infty$,
then the time average
$\overline{x}(t) =\int_0 ^t x(t') {\rm d} t'/t$ is equal to
the corresponding ensemble average $\langle x \rangle_{{\rm eq}}$.
In experiment the measurement time might be long, but it is
always finite. Hence it is natural to ask what  the fluctuations
of $\overline{x}$ are. Such an analysis 
sheds light on deviations from the thermal
 equilibrium average
 due to finite time
measurement, a general theme which has attracted much interest in the context
of fluctuation theorems \cite{Fluc}. 
The
Boltzmann measure, due to ergodicity, yields equilibrium properties 
of thermal  systems. Surprisingly, we find
that for Langevin dynamics, the Boltzmann measure also determines the
deviations from ergodicity.

\begin{figure}
\centering
\epsfig{figure=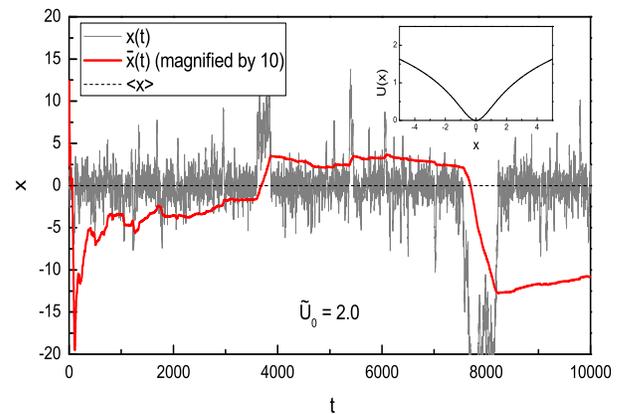, totalheight=0.23\textheight, width=0.47\textwidth, trim=25mm 5mm 25mm 15mm, clip}
\caption{ 
(color online) The trajectory of a Brownian particle in a logarithmic potential
exhibits large amplitude fluctuations. As a consequence,
the time average of the process 
$\overline{x}$ (red curve) does not converge to a fixed value even though Boltzmann
equilibrium ensemble average $\langle x \rangle_{{\rm eq}}$ is zero. 
Here 
$U(x)= \ln(1 + x^2)/2$,
$k_B T = 1/2$ and the diffusion constant $D=1$. 
}
\label{fig1}
\end{figure}

 As we will show, for binding fields $U(x)$ where the Fokker-Planck (FP)
operator exhibits a discrete 
eigenspectrum, the fluctuations of the time average
$\overline{x}$ become small as time increases, as expected from
ordinary ergodic statistical mechanics. 
 For this type of field, ergodicity is related to 
the work of Szabo, Schulten, and Schulten \cite{Szabo} on the seemingly unrelated problem 
of reaction
kinetics (see details below). 
 A more interesting case is that of a
 logarithmic
binding field \cite{Lutz}  $U(x) \sim U_0 \ln(|x|)$ when $|x| \to \infty$,
 since for such a potential the
fluctuations of $\overline{x}$ are not small even in the long time limit. 
Here the Boltzmann measure exhibits power law tails, $P^{\rm eq} (x) \propto
|x|^{ - U_0/(k_B T)}$. Starting at the origin, the particle during its evolution
tends to sample larger and larger values of $|x|$ as illustrated in
Fig. \ref{fig1}. 
Large fluctuations in the amplitude of $x(t)$ cause the time average
of this special process to remain random even in the long time
limit. In what follows, we calculate the magnitude of these fluctuations
and show how they are related to a super-aging correlation function.
Importantly, such logarithmic potentials model many physical systems,
ranging from optical lattices \cite{Zoller}, 
charges in vicinity of a long
charged polymer \cite{Manning}, 
DNA dynamics \cite{Fogedby}, membrane induced forces \cite{Farago},
a nano-particle in a trap \cite{Adam}, 
to long ranged interacting models \cite{Chavanis}.
At the end of this Letter we discuss the connection between
our theory and a recent experiment \cite{Sagi}.

{\em Model and  observable.} 
Brownian dynamics in a force field 
$f(x)= - {\rm d}  U(x)/ {\rm d} x$ obeys the equation \cite{Risken}
\begin{equation}
{{\rm  d} x \over {\rm d} t } =  - { f(x) \over   \gamma} + \eta(t).
\label{eq01}
\end{equation}
Here $\gamma$ is the friction constant, 
$\eta(t)$ is Gaussian white noise obeying the
fluctuation dissipation relation
$\langle \eta(t) \eta(t')\rangle = 2 D \delta(t-t')$,  and 
$D=k_B T/ \gamma$ according to the Einstein relation.
From the trajectory 
$x(t)$ we construct the 
time average $\overline{x}(t) = \int_0 ^t x(t') {\rm d} t' /t$. 
For a binding potential,  in the long time limit, $x$ obeys the equilibrium Boltzmann  distribution:
\begin{equation}
P^{{\rm eq}} (x) = { \exp\left[ - {U(x) \over k_B T} \right] \over Z}; \quad Z=\int_{-\infty} ^\infty e^{-\frac{U(x)}{k_B T}} {\rm d} x
\label{eq02}
\end{equation} 
where $Z$
is the normalizing partition function {\em which is assumed to be finite}.
We consider symmetric potentials $U(x) = U(-x)$ and then
the ensemble average in equilibrium $\langle x \rangle_{{\rm eq}}= 
\int_{-\infty} ^\infty x P^{{\rm eq}} (x) {\rm d} x =0 $.  
If the process is ergodic then in the long time limit 
$\overline{x} \to \langle x \rangle_{{\rm eq}}=0$. If 
 $\lim_{t \to \infty} \langle \overline{x}^2(t) \rangle \neq 0$ the process is
non-ergodic, where $\langle \cdots \rangle$
stands for an ensemble mean. 
 In the second
part of our work we show that not all binding potentials satisfy 
the ergodic hypothesis. 

{\em Szabo-Schulten-Schulten equation yields the fluctuations
of the time average.} The variance of 
the time average is given
by
\begin{equation}
\langle \overline{x}^2(t) \rangle = {1 \over t^2} \int_0 ^t {\rm d } t_2 \int_0 ^t {\rm d} t_1 \langle x(t_2) x(t_1) \rangle
\label{eq03}
\end{equation} 
where $\langle x(t_2) x(t_1) \rangle$ is the correlation function. 
For the Markovian process under investigation, and for a particle starting at
the origin at time $t=0$ we have \cite{Risken}
\begin{widetext}
\begin{equation}
\langle \overline{x}^2(t) \rangle = {2 \over t^2} \int_0 ^t {\rm d } t_2 \int_0 ^{t_2 } {\rm d} t_1 \int_{-\infty} ^\infty \int_{-\infty} ^\infty x_2 x_1 P\left(x_2, t_2| x_1 , t_1\right) P\left( x_1, t_1 | 0, 0\right){\rm d} x_1 {\rm d} x_2 
\label{eq04}
\end{equation} 
\end{widetext}
where $P(x_2,t_2|x_1,t_1)$ is the conditional probability density to find
the particle on $x_2$ at time $t_2$ once it is located at $x_1$ at time
$t_1$. 
In the limit of long times, 
the major contribution to the integration over $t_1$ comes from
long times; hence one replaces $P(x_1,t_1|0,0)$ with 
$P^{{\rm eq}} (x_1)$. To proceed, it is useful to define
\begin{equation}
\xi(x_1) = \int_0 ^\infty E(x_1,\tau){\rm d} \tau 
\label{eq05} 
\end{equation} 
where $E(x_1,\tau)$ is the averaged position of
a particle at a time $\tau$  after it starts at $x_1$.  
Two cases are of interest; the first is when $\xi(x_1)$ is finite,
the other when it diverges. We shall start with the former case
which is clearly relevant to potential fields where the 
 FP eigenspectrum \cite{Risken} has a finite energy gap to the ground state, 
 since then the relaxation of $E(x_1,\tau)$
is exponential.
From Eq. (\ref{eq04}) it follows that in the long time
limit
\begin{equation} 
\langle \overline{x}^2(t) \rangle \sim {2 \over t} \int_{-\infty}
 ^\infty  x_1 \xi(x_1) P^{{\rm eq}} (x_1){\rm d} x_1 .
\label{eq06}
\end{equation} 
As is well known the backward FP equation 
\cite{Risken}
\begin{equation}
\begin{array}{l}
L^{\dagger} _{{\rm FP}} P(x_2, \tau|x_1,0) = {\partial \over \partial \tau} P(x_2,\tau|x_1,0), \\
\ \ \\
 L^{\dagger} _{{\rm FP}} = D \left[ {\partial^2 \over \partial (x_1)^2 } + { f(x_1) \over k_B T}  {\partial \over \partial x_1} \right]
\end{array}
\label{eq07}
\end{equation} 
governs the dynamics
where $L^{\dagger} _{{\rm FP}}$ is
the adjoint FP operator and $P(x_2,0|x_1,0)=\delta(x_2 - x_1)$.
By definition
$E(x_1,\tau)=\int_{-\infty} ^\infty x_2 P(x_2,\tau|x_1,0){\rm d} x_2 $
which implies
\begin{equation}
L^{\dagger} _{{\rm FP}} E(x_1,\tau)  = 
{\partial \over \partial \tau} E(x_1,\tau) 
\label{eq11}
\end{equation} 
with $E(x_1,0) = x_1$.
Using Eq. (\ref{eq05}), we find
\begin{equation} 
L^{\dagger} _{{\rm FP}} \xi(x_1) = - x_1
\label{eq08}
\end{equation} 
with $\xi(0)=0$. 
Eq. (\ref{eq08}) was obtained previously
in \cite{Szabo}  in the context of reaction
kinetics. 
Eqs. (\ref{eq04}-\ref{eq08}) are so general that they could be extended 
to arbitrary  Markovian processes. 
The latter equations thus serve as a starting point
for the investigation of fluctuations of time averages for a wide class
of systems. 

{\em Fluctuations of time averages determined from  Boltzmann statistics.}
Eq. 
(\ref{eq08})  
is easy to solve,
and upon using Eq. (\ref{eq06}) we find the general formula
\begin{equation}
\langle \overline{x}^2 \rangle \sim { 2 \over  D t} \int_{-\infty} ^\infty {e^{U(x) /  (k_B T)} \over Z} {\rm d} x  \left[ \int_{x}  ^\infty x' e^{ - U(x')/ (k_B T)} {\rm d} x' \right]^2 . 
\label{eq09}
\end{equation}
As is well known, Boltzmann statistics can be used
to determine the time average of  ergodic processes:
$\overline{x} \to \langle x \rangle$ in the long time limit.
Eq. (\ref{eq09}) shows that  also 
the finite time fluctuations of  $\overline{x}$ are determined by the
Boltzmann distribution.
Surprisingly, Eq. (\ref{eq09}) shows that the difficult task of finding
the entire eigenspectrum of the FP operator is not required. 
Eq.
(\ref{eq09}) is easily generalized to dimensions greater than one, and
to non-thermal processes whose equilibrium density is non-Boltzmannian. 
As expected from ergodicity, the magnitude of the  fluctuations  
decays to zero with time, provided that
the integrals in Eq. (\ref{eq09})  converge. For example,
for the harmonic potential $U(x) = m \omega^2 x^2 /2$ we get
$\langle \overline{x}^2 \rangle \sim 2 (k_B T)^2 / [ D ( m \omega^2)^2 t]$. 
An interesting case  where the integrals diverge
is the  logarithmic potential 
$U(x) \sim U_0  \ln(|x|) $ for $|x| \to \infty$ and $U_0/(k_B T) < 5$.
This leads to a  non-ergodic behavior which we now investigate.

{\em Logarithmic potential.}  
We will first find the  two-point
correlation  function $\langle x(t_2) x(t_1) \rangle$
for a general logarithmic  
potential which satisfies $U(x) \sim U_0 \ln(|x/a|)$, e.g.
 $U(x) = 0.5 U_0 \ln[1+ (x/a)^2]$.
We will then use (\ref{eq03}) to obtain the fluctuations of the time average
showing that for high enough temperature the fluctuations 
increase with time. 
For this potential, for $1<U_0/(k_B T)<5$, due to the slow convergence of the tail of the distribution
to $P^{\rm eq}$, and 
the slow power-law decay of $E(x_1,\tau)$ (which we shall shortly demonstrate), rendering $\xi(x_1)$ infinite for $U_0/(k_B T)<2$, one must consider the
full time dependent problem instead of  the time independent
 Eq. (\ref{eq08}) and  $P^{{\rm eq}}(x)$. 
Generally the correlation function is given by
\begin{equation}
\langle x(t_2) x(t_1 ) \rangle= \int_{-\infty} ^\infty x_1 E(x_1, t_2 - t_1) P(x_1, t_1|0,0){\rm d} x_1 .
\label{eq10}
\end{equation} 
To solve this problem we used two approaches; the first
is based on an eigenfunction expansion of the
solution of the FP equation \cite{Andreas}.
 Such a calculation
is lengthy and hence we adopt here a scaling approach.
As seen from Eq. (\ref{eq10})
the key quantity to calculate is the ensemble mean 
$E(x_1,\tau)$ using Eq. 
(\ref{eq11}).
Due to the homogenous character of the large $x$ Fokker-Planck operator, 
it is natural to adopt a scaling ansatz:
\begin{equation}
E(x_1,\tau)\sim \tau^\alpha g\left( { x_1 \over \tau^\beta} \right) 
\label{eq13}
\end{equation} 
where $\alpha$ and $\beta$ are scaling exponents. Since for short time
$E(x_1,\tau)\simeq x_1$
we have $g(y) \simeq y$ for large $y$ 
and $\alpha=\beta$. 
Inserting Eq. (\ref{eq13}) in Eq. (\ref{eq11})
we find to leading order 
\begin{equation}
\tau^{-\beta} D \left( g''- \tilde{U}_0 {g'\over y}\right) = \tau^{\beta- 1} \beta \left( g - y g'\right),
\label{eq14}
\end{equation} 
where $\tilde{U}_0 = U_0 /(k_B T)$ is a key dimensionless parameter. 
To achieve a $t$-independent equation,
we must have $\beta=1/2$, 
 typical of Brownian
motion. 
Then,
\begin{equation}
g(y) = c_1 y^{1+ \tilde{U}_0} e^{ - {y^2 \over  4 D}} M\left({3 \over 2} , {3 + \tilde{U}_0 \over 2} , {y^2 \over 4 D} \right)
\label{eq15} 
\end{equation} 
where $M(a,b,x)$ [also denoted $_1 F_1 (a;b;x)$]
 is the  Kummer $M$ function~\cite{Abr}  and
we rejected a second solution in terms of the Kummer $U$ function
 since
it does not satisfy the boundary condition 
$E(x_1,\tau)  \to 0$ when $\tau \to \infty$ 
(i.e., relaxation to equilibrium).
The constant $c_1$ is found by matching the solution
in the $y\to \infty$ limit which corresponds to short times.
Using $M(a,b,x)   \sim  \exp(x) \Gamma(b) x^{a-b}/\Gamma(a)$ and
$g(y) \sim y $  we find
$c_1 = \{\Gamma(3/2) / \Gamma[(3 + \tilde{U}_0 )/2]\}(4 D)^{- \tilde{U}_0 /2}$. 
In particular, for long times, $E(x_1,\tau) \sim \tau^{-\tilde{U}_0/2}$,
 so as we claimed,
$\xi(x_1)$ diverges for $\tilde{U}_0<2$.

{\em Steady state cannot be used to obtain the correlation function.}
To complete the calculation, we must have $P(x_1,t_1|0,0)$ which 
was recently obtained \cite{KesslerPRL}.
 The equilibrium PDF, since it decays as a power law  
$P^{{\rm eq}} (x) \propto |x|^{- \tilde{U}_0} $ 
%
would give, for $1<\tilde{U}_0<3$,
 $\langle x(t_2) x(t_1) \rangle = \infty$ for
$t_1=t_2$. This is an unphysical behavior:  at finite time
 one cannot have an infinite value for the correlation function, 
 since the particle cannot
travel faster than diffusion permits. 
Specifically in the limit
of long $t_1$ we have  \cite{KesslerPRL}
\begin{equation}
P(x_1,t_1| 0 ,0 ) \sim P^{{\rm eq }} (x_1) {\Gamma( { 1 + \tilde{U}_0 \over 2} , {(x_1)^2 \over  4 D t_1}  ) \over \Gamma( { 1 + \tilde{U}_0 \over 2} )}. 
\label{eq16} 
\end{equation} 
Since $\Gamma(a,0)=\Gamma(a)$ as  $t \to \infty$, 
thermal equilibrium is reached. 
Nevertheless, for the calculation
of correlation functions one must take into account the finite time correction
which is represented by the ratio of $\Gamma$ functions.  

\begin{figure}
\centering
\epsfig{figure=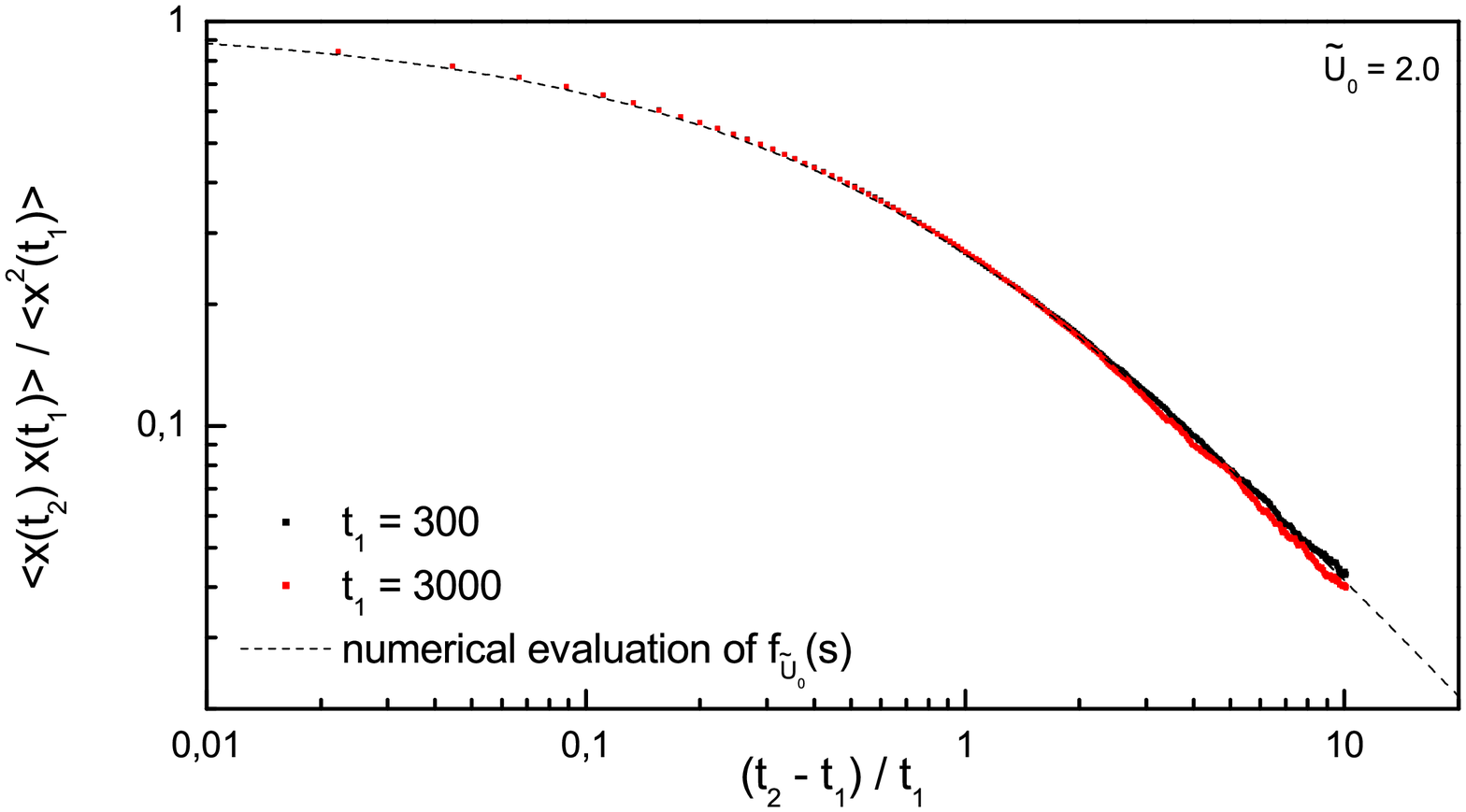, totalheight=0.23\textheight, width=0.47\textwidth, trim=18mm 10mm 25mm 15mm, clip}
\caption{ 
The aging correlation function 
Eq. (\ref{eq18})
perfectly matches numerical simulation of the Langevin Eq.
(\ref{eq01}). 
}
\label{fig2}
\end{figure}

{\em Aging correlation function.} Inserting Eqs.
 (\ref{eq15},\ref{eq16}) in Eq. (\ref{eq10}) 
we find the non-stationary correlation function for
the temperature range $1 < \tilde{U}_0 < 3$: 
\begin{equation}
\langle x(t_2) x(t_1) \rangle \sim \langle x^2 (t_1) \rangle f_{\tilde{U}_0}\left( { t_2 - t_1 \over t_1} \right)
\label{eq18}
\end{equation}
where
\begin{equation}
\begin{array}{c}
f_{\tilde{U}_0} (s) =
 {  \sqrt{ \pi} \left(3 - \tilde{U}_0\right) \over 2 \Gamma \left( { 3 + \tilde{U} _0 \over 2} \right) } s^{ {3 - \tilde{U}_0 \over 2}} \times \\ 
\int_0 ^\infty {\rm d} y y^2 e^{- y^2} M\left( { 3 \over 2} , {3 + \tilde{U}_0 \over 2} , y^2 \right) \Gamma \left( { \tilde{U}_0 + 1 \over 2}, y^2 s \right).
\end{array}
\label{eq19}
\end{equation}
The behavior in Eq. (\ref{eq18}) is very different than the stationary case
where the correlation function is a function of the time difference
$t_2 - t_1$. 
In  this temperature 
 regime the equilibrium mean square displacement diverges, 
$\langle x^2 \rangle_{{\rm eq}}= \infty$, while the
time dependent solution 
Eq. (\ref{eq16}) 
gives~\cite{KesslerPRL}
$ \langle x^2 (t_1) \rangle = a^2 c_2 (4  D t_1/ a^2)^{ (3 - \tilde{U}_0 )/2}$, 
$c_2=2 (a/Z) [\Gamma(1/2+\tilde{U}_0/2)(3 - \tilde{U}_0)]^{-1}$.
We  find $f_{\tilde{U}_0} (0) = 1$ which implies that 
$C(t_1,t_1) = \langle x^2(t_1) \rangle$ as it should. 
In the opposite limit $t_2 \gg t_1$, we obtain 
\begin{equation}
 \langle x(t_2) x(t_1) \rangle \sim  c_3 \langle x^2  (t_1) \rangle \left( { t_2 \over t_1} \right)^{ - \tilde{U}_0 \over 2} 
\label{eq20}
\end{equation}
with $c_3=(3/2-\tilde{U}_0/2)\sqrt{\pi}\Gamma(2+\tilde{U}_0/2)/3 \Gamma(3/2+\tilde{U}_0/2)$.
In Fig. \ref{fig2} we compare our analytical 
Eq. (\ref{eq19}) with Langevin simulations
showing excellent agreement for various measurement times. 

 As mentioned we assume that the partition function function $Z$ is finite
and hence the steady state $P^{{\rm eq}}(x)$ is normalizable. 
This 
 excludes the well known Bessel process \cite{Bray}
which can be mapped onto $U(x)= U_0  \ln|x|$ with its singularity 
at the origin. 
It is important to emphasize
that $\langle x(t_2) x(t_1) \rangle \sim 1/Z$  depends on the shape of
the  potential
{\em in the whole space} through $Z$.   
Hence for the calculation of the correlation function
the regularity of the potential on the origin
is vital. Interestingly this is not the case for all observables; e.g., 
$E(x_1,\tau)$ 
Eqs. (\ref{eq13},\ref{eq15})
is $Z$ independent and hence related to the Bessel process \cite{Bray}. 

\begin{figure}
\centering
\epsfig{figure=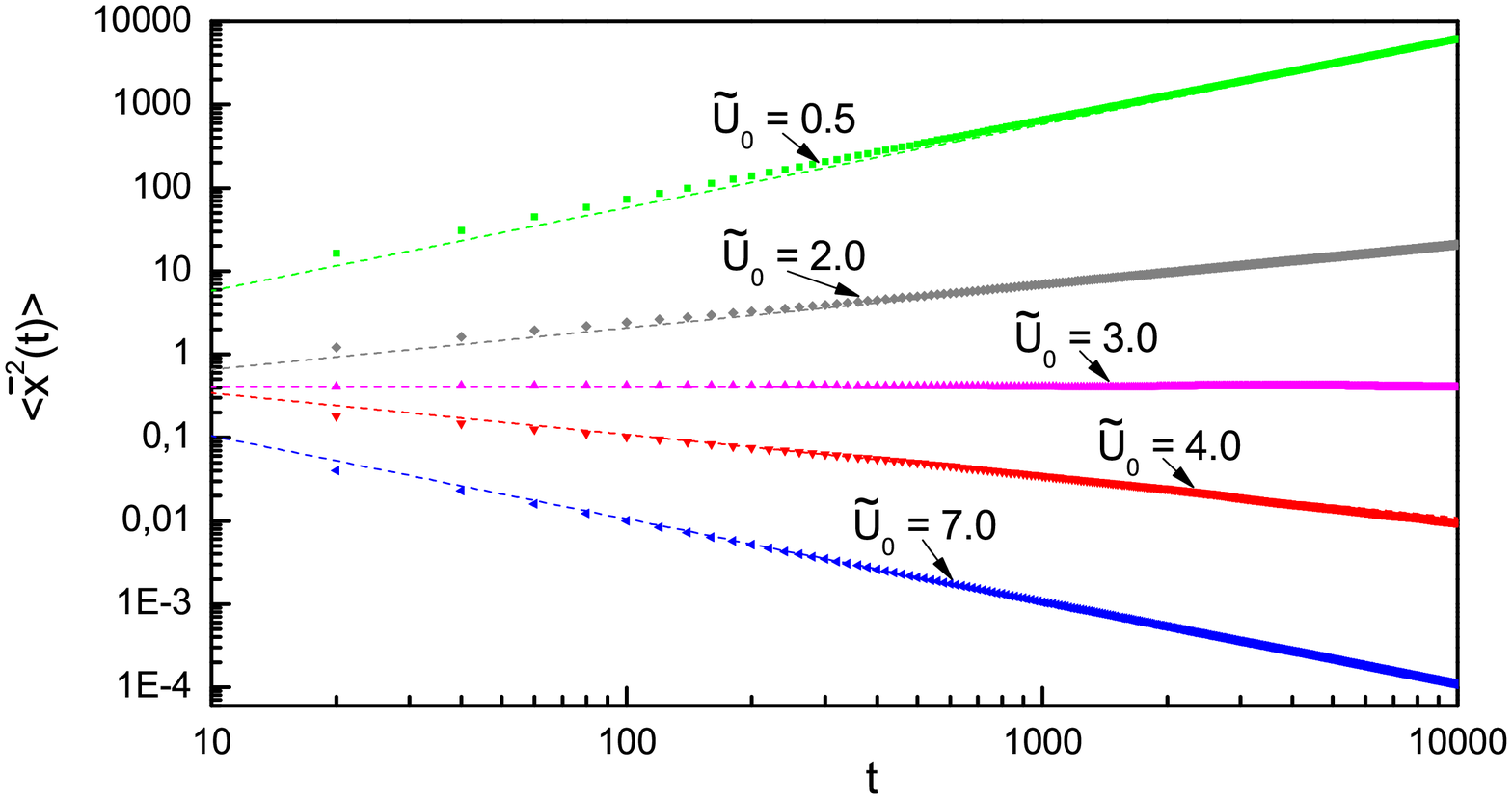, totalheight=0.23\textheight, width=0.47\textwidth, trim=18mm 10mm 25mm 15mm, clip}
\caption{ 
 In
the non-ergodic phase 
 $k_B T> U_0/3$ ($\tilde{U}_0 < 3$), 
$\langle \overline{x}^2(t) \rangle$ increases with time.
For the critical point $k_B T=U_0/3$ ($\tilde{U}_0 = 3$) the fluctuations are constant.
The dashed curves are theoretical predictions Eqs. (\ref{eq22},\ref{eq23})
 which agree 
very well with the numerical simulations.  
}
\label{fig3}
\end{figure}

{\em Ergodicity of the
dynamics} is classified in four domains which are controlled by 
temperature.\\ 
{\bf (a)} The most interesting case is the regime $1<\tilde{U}_0 <3$.
As we showed, a normalized steady state exists and from symmetry
$\langle x \rangle_{{\rm eq}} = 0$. If we naively assume ergodicity
$\overline{x} \to \langle x \rangle_{{\rm eq}}=0$ and
$\lim_{t \to \infty} \langle \overline{x}^2(t) \rangle=0$.
 Rather, from  Eqs.    
(\ref{eq03},\ref{eq18}) we find \cite{remark}
\begin{equation}
\langle \overline{x}^2(t) \rangle \sim { 2 \langle x^2(t) \rangle  \over t^2} \int_0 ^t {\rm d} t_1 \int_{t_1} ^t {\rm d} t_2 \left( { t_1 \over t } \right)^{{ 3 - \tilde{U}_0 \over 2} } f_{\tilde{U}_0 } \left( { t_2 - t_1 \over t_1} \right).
\label{eq21}
\end{equation}
Changing variables to  $s=t_2/t_1-1,w=t/t_1-1$,
we find 
\begin{equation}
\langle \overline{x}^2(t) \rangle \sim c_4 \langle x^2(t) \rangle  \propto t^{3 - \tilde{U}_0 \over 2} 
\label{eq22}
\end{equation}
where
$c_4=4\int_0 ^\infty {\rm d} w (1 + w)^{(\tilde{U}_0 - 7)/ 2}  (7-\tilde{U}_0)^{-1} f_{\tilde{U}_0} (w)$.
We see that the fluctuations grow with time, hence 
ergodicity is broken. 
We find that $c_4 \approx 0.2397$ for $\tilde{U}_0 = 1$ and 
that it decreases monotonically to $c_4 = 0$ at $\tilde{U}_0 = 3$.\\
{\bf (b)} For lower temperature, $3 < \tilde{U}_0< 5$, the integrals in Eq. (\ref{eq09}) still diverge, and   $\langle \overline{x}^2(t) \rangle$ decays as
$t^{(3 - \tilde{U}_0) / 2} $;  indicating   an anomalously
slow approach to ergodicity.
%
%
 \\
{\bf (c)} For $\tilde{U}_0 >5$ the temperature is low enough that
Eq. (\ref{eq09}) is now valid.
 For $U(x) = 0.5 U_0 \ln[1+ (x/a)^2]$ we find
\begin{equation}
\langle \overline{x}^2 (t)\rangle \sim { 2 (\tilde{U}_0 -4) \over 
(\tilde{U}_0 - 2) (\tilde{U}_0 - 3) } { a^4 \over (\tilde{U}_0 - 5) D t} 
\label{eq23}
\end{equation}
which diverges when $\tilde{U}_0 \to 5$. \\
{\bf (d)} Finally, for very high temperatures $\tilde{U}_0 < 1$, the equilibrium state Eq.
(\ref{eq02}) 
is not  defined as the partition function $Z$
diverges. Here $\langle \overline{x}^2(t) \rangle \propto t$, 
exactly the diffusive behavior of a free particle, $U_0=0$~\cite{Andreas}. \\
These four different behaviors are confirmed via
numerical simulations presented in Fig.
\ref{fig3}, which illustrates convergence on
reasonable computer time scales. 
A summary of the scaling regimes is presented in Table 1. 

\begin{table}
\begin{tabular}{|c | c | c|}
\hline
 \ &\  $\langle \overline{x}^2(t) \rangle$ \ & \ $\langle x^2(t) \rangle$ \ \\
\hline
$\tilde{U}_0 < 1$\  &\ $t$ \  &\ $t$ \ \\
\hline
$1 <\tilde{U}_0 < 3\ $ &\ $t^{(3 - \tilde{U}_0)/2}$ \  & \ $t^{(3 - \tilde{U}_0)/2}$ \ \\
\hline
$3 <\tilde{U}_0 < 5$ & $t^{(3 - \tilde{U}_0)/2}$   & $t^0$ \\
\hline
$5 <\tilde{U}_0 $ &$ t^{-1}$   & $t^0$ \\
\hline
\end{tabular}
\caption{Scaling behavior of $\langle \overline{x}^2 (t) \rangle$ and
$\langle x^2 (t) \rangle$ for various values of $\tilde{U}_0 = U_0/(k_B T)$. }
\label{Tab1}
\end{table}

{\em Relation with experiment.}
After the submission of this manuscript, an experiment
on anomalous diffusion of
ultra-cold atoms which employs the well known Sisyphus
cooling scheme was reported \cite{Sagi}. 
In the semi-classical approximation, the atomic velocity distribution
follows Fokker-Planck dynamics in an asymptotically logarithmic potential
\cite{Zoller,KesslerPRL,Renzoni}.
Our work provides the theoretical mean-square displacement in
this experiment by identifying our position $x$ with the velocity 
$v$ of the atoms. The measured atomic position
is $x(t) = \int_0 ^t v(t) {\rm d} t$ and hence $x(t)/t$ corresponds to the
time averaged velocity. The PDF of the atoms in the experiment has been
described with a L\'evy distribution, with a divergent variance.
However, our results show that the mean square displacement is finite
for any finite measurement time. 
These seemingly contradicting findings 
are related to the well known dilemma whether L\'evy flights
are at all physical, since they predict diverging mean square displacement,
which must be tamed  
\cite{Zoller,Klafter}.  
We speculate that the L\'evy distribution found in the experiment describes
the center part of the packet, which eventually is cut off to give a
finite mean square displacement. 
Furthermore, using our results one can estimate
the time in which the atoms remain within a finite domain, which is
of course crucial for experiments. Experimentally one may also control
the depth of the optical potential, here modeled with $\tilde{U}_0$ and hence
explore the nontrivial dependence of our results on this parameter.  
We will elaborate on these interesting points in a longer publication.

{\em Discussion.} Aging correlation functions and ergodicity breaking
typically describe glassy dynamics  
\cite{Fisher,PNAS} (and Ref. therein).
Our work shows that aging and ergodicity breaking
can be found also for simple Markovian dynamics,
 without the need to introduce
heavy-tailed waiting times into the kinetic scheme,
 nor disorder or  many-body physics. 
The aging correlation function
(\ref{eq18})
has a signature very different than most previous work.  The
prefactor  
$\langle x^2 (t_1) \rangle$ grows with time,
and hence we call it super-aging. 
This is in contrast to normal aging where the correlation
function is of the form
$C(t_2,t_1) = \langle x^2 \rangle_{{\rm eq}}f(t_2/t_1)$ 
with a finite equilibrium value $\langle x^2 \rangle_{{\rm eq}}$. 
A similar non-normal aging behavior, albeit with a logarithmic
time dependence, has been found in Sinai's model of diffusion in
a random environment \cite{Fisher}. 
Unlike  previous scenarios to ergodicity breaking,
the amplitude of the stochastic process
$x(t)$  in our work increases with time, since the particle
explores more and more of the tails of the equilibrium PDF as time goes on. 
Thus rare events where the amplitude $x(t)$
of the Markovian process attains a large
value are responsible for the non-ergodic behavior. 
 This is clearly related to the power law tail of the
equilibrium steady state 
$P^{{\rm eq}} (x) \propto |x|^{- \tilde{U}_0}$.
More importantly, physical systems with fat tailed equilibrium states
are common and hence this type of ergodicity breaking
may find broad applications.

{\bf Acknowledgement} 
This  work  was supported by the  Israel Science  Foundation, 
the Emmy Noether Program of the DFG (contract No LU1382/1-1) and the
cluster of excellence Nanosystems Initiative Munich.


\begin{thebibliography}{99}

\bibitem{Fluc} C. Jarzynski, {\em  Annu. Rev. Condens. Matter Phys.} {\bf 2}, 329 (2011).

\bibitem{Szabo} A. Szabo, K. Schulten,  and Z. Schulten, {\em J. Chem. Phys.} 
{\bf 72}, 4350 (1980).

\bibitem{Lutz} E. Lutz, {\em Phys. Rev. Lett.} {\bf 93}, 190602 (2004).

\bibitem{Zoller} S. Marksteiner, K. Ellinger, and P. Zoller,
{\em Phys. Rev. A} {\bf 53}, 3409 (1996).


\bibitem{Manning} G. S. Manning, {\em J. of Chemical Physics} {\bf 51}, 924 (1969).

\bibitem{Fogedby} H. C. Fogedby, and R. Metzler, 
{\em Phys. Rev. Lett.} {\bf  98}, 070601 (2007). 

\bibitem{Farago} O. Farago, {\em Phys. Rev. E} {\bf 81}, 050902 (2010). 

\bibitem{Adam} A. E. Cohen, {\em Phys. Rev. Lett.} {\bf 94}, 118102 (2005).

\bibitem{Chavanis} P. H. Chavanis  and  R. Mannella, 
{\em Eur. Phys. J. B}  {\bf 78}, 139 (2010).

\bibitem{Sagi} Y. Sagi, M. Brook, I. Almog, and N. Davidson
arXiv:1109.1503v1 [quant-ph] (2011). 

\bibitem{Risken} H. Risken, {\em The Fokker Planck Equation} Springer 
1996 (Berlin).




\bibitem{Bray} A. J. Bray, {\em Phys. Rev. E} {\bf 62}, 103 (2000).


\bibitem{Andreas}  A. Dechant, Diploma thesis, University of Augsburg (2011). 

\bibitem{Abr} M. Abramowitz and I. A. Stegun, {\em Handbook of Mathematical Functions} Dover 1972 (New York) 

\bibitem{KesslerPRL}  D.A. Kessler and E. Barkai, {\em Phys. Rev. Lett.} {\bf 105}, 120602 (2010).

\bibitem{remark} This 
 is related to the observation that the FP eigenspectrum 
for the logarithmic potential
has no gap. A. Dechant, E. Lutz, E. Barkai, D. A. Kessler
{\em J. Stat. Phys.} {\bf 145}, 1524 (2011).
 

\bibitem{Renzoni} P. Douglas, S. Bergamini, and F. Renzoni {\em Phys. Rev. Lett.} {\bf 96}, 110601 (2006). 

\bibitem{Klafter} J. Klafter, M. F. Shlesinger, and G. Zumofen
{\em Physics Today} {\bf 49} 33 (1996). 

\bibitem{Fisher} D. S. Fisher, P. Le Doussal, and C. Monthus,
{\em Phys. Rev. Lett.}{\bf 80}, 3539 (1998).
 P. Le Doussal, C. Monthus and D. S. Fisher
{\em Phys. Rev. E} {\bf 59} 4795 (1999). See Eq. (159) therein.

\bibitem{PNAS} 
J. P. Bouchaud, {\em J. Phys. I} France {\bf 2}, 1705 (1992). 
S. Burov, R. Metzler, E. Barkai,
{\em PNAS} {\bf 107}, 13228 (2010).


\end{thebibliography}
\end{document}